# Comparing Methodologies for Ranking Alternatives: A case study in assessing bank financial performance

Dong Trung Chinh[1], Nguyen Thi Thu Hien[2], Pham Huong Quynh[2], Vu Quang Minh[3*]
[1]Hanoi College of Industrial Economics
[2]. Hanoi Univesity of industry
[3]Northern Kentucky University

**Abstract**

Bank financial performance encapsulates an institution's capacity to effectively manage its assets, capital, and operational activities to generate profits and ensure stability. Evaluating this performance necessitates the integration of diverse metrics, including profitability indicators, loan growth rates, capital utilization efficiency, and more. Nevertheless, directly comparing the financial performance across different banks presents a complex challenge due to inherent disparities in their specific performance parameters. Multi-criteria decision-making (MCDM) techniques are frequently employed to navigate this intricate assessment. This study undertakes a comparative analysis of various MCDM approaches in evaluating bank financial performance. Our investigation encompasses both a comparison of methods for assigning weights to criteria and a comparison of methodologies for ranking the alternatives (banks). We examine five distinct weighting methods: Equal, Entropy, MEREC, LOPCOW, and SPC. Concurrently, three alternative ranking methods Probability, TOPSIS, and RAM are compared. These comparisons are conducted within the context of a case study involving the performance assessment of 19 banks. The findings indicate that the highest degree of stability in ranking bank financial performance is achieved when the Entropy method is utilized for criteria weighting in conjunction with the Probability method for ranking alternatives.

**Keywords**: bank financial performance, weighting, MCDM

## 1. Introduction

Bank financial performance represents a comprehensive metric reflecting the efficacy with which capital and assets are deployed to generate profits and maintain operational stability. It extends beyond mere figures on financial statements, encompassing a holistic view of risk management capabilities, resource optimization, and contributions to economic development. Comparing the financial performance across banks is not merely a statistical exercise; it holds profound strategic implications for various stakeholders.

Assessing financial performance empowers banks to pinpoint strengths and weaknesses, facilitating strategic adjustments and operational improvements to enhance competitiveness and foster sustainable growth. For the public and investors, transparent performance data informs judicious decisions regarding deposits, loans, or stock investments, guiding them toward security and profitable opportunities [1,2]. Businesses, in turn, leverage bank financial performance evaluations to select appropriate financial partners aligned with their needs, gaining access to more competitive products, services, and interest rates [3, 4]. Regulatory bodies and policymakers utilize these assessments to gauge systemic health, identify emerging risks, and subsequently formulate operational policies and legal frameworks. This ensures the stability of the national financial system and safeguards depositors' interests [3].

Evaluating bank financial performance necessitates reliance on multiple factors, such as profitability, asset quality, and liquidity. However, these parameters vary significantly across individual banks, rendering the comprehensive assessment of bank financial performance a complex undertaking [5, 6]. Multi-criteria decision-making (MCDM) is recognized as a valuable tool widely applied in the realm of bank evaluation [7, 8]. Nevertheless, the ranking of alternatives is considerably influenced by factors like the specific MCDM method (ranking method) employed and the criteria weighting technique utilized [9-12]. Therefore, it's crucial to conduct comparative analyses among MCDM methods and among weighting methods to identify approaches that yield highly stable sets of alternative rankings.

Among the myriad MCDM methods, TOPSIS is widely acknowledged as the most commonly used [13, 14]. Probability, conversely, stands out as a distinct method that, by eschewing "additive" calculations, ensures high reliability in its generated results [15, 16]. RAM is a relatively recent addition to the field, offering the advantages of simple operations, fewer computational steps, and the ability to balance between benefit-type and cost-type criteria [17, 18]. The characteristics of TOPSIS, Probability, and RAM, as just outlined, prompt a crucial question: Which method proves most suitable for evaluating bank financial performance? The pursuit of an answer to this question motivates the concurrent application of all three methods in this research.

The majority of MCDM methods, including TOPSIS, Probability, and RAM, mandate the calculation of criteria weights. These weights profoundly influence the rankings of the alternatives under consideration [19]. An alternative might rank favorably with one weighting method but poorly with another [20]. This presents a challenge: how does one select a criteria weighting method that ensures rank stability among alternatives? The five objective weighting methods examined Equal, Entropy, MEREC, LOPCOW, and SPC derive their weights solely from the intrinsic values of the parameters within each alternative, independent of subjective human judgment [21]. Equal is the simplest, assigning uniform weights to all criteria. Entropy and MEREC are widely adopted and recommended in numerous studies, including recent ones [22, 23]. LOPCOW [24] and SPC [25] are newer methods that have emerged recently. All five of these methods have been applied to criteria weighting across diverse fields. The central question here is which of these five methods is best suited for weighting criteria when evaluating bank financial performance. The exploration of this question underpins the inclusion of all five methods in this study.

In summary, the primary objective of this research is to compare the TOPSIS, Probability, and RAM methods, as well as the Equal, Entropy, MEREC, LOPCOW, and SPC weighting methods, in the context of evaluating bank financial performance. This comparison aims to identify the most suitable methods for practical application. Section 2 presents a concise literature review of existing works that apply MCDM methods to bank evaluation. Section 3 outlines the dataset pertaining to the banks under financial performance evaluation, sourced from a recently published document. This section also briefly details the application steps for TOPSIS, Probability, RAM, Equal, Entropy, MEREC, LOPCOW, and SPC, along with the criteria used for method comparison. The results of the method comparisons are presented in Section 4. Finally, the conclusions drawn and directions for future research are discussed in the concluding section of this paper.

**2. Literature review**

In the nascent stages of applying Multi-Criteria Decision-Making (MCDM) methods to bank evaluation, many studies exclusively employed a single MCDM technique for specific problems. A select few of these are listed below. The TOPSIS method, for instance, was utilized to rank eight commercial banks in Vietnam, with criteria weights determined by the evaluators' subjective opinions [2]. Similarly, [26] exclusively applied TOPSIS to assess banks in China. TOPSIS was also employed to evaluate six Indian banks, with criteria weights computed via the AHP method [27], and to assess twelve banks in Turkey [28]. In another study, the PROMETHEE method was used to evaluate five Indian banks [29]. Clearly, the sole reliance on a single MCDM method for ranking banks represents a significant limitation in such research, as the introduction highlighted that alternative rankings can shift considerably when different methods are employed [9, 10].

The constraint of using a single MCDM method for each specific problem has since been addressed through the concurrent application of multiple methods in particular cases. Some of these studies have demonstrated consistent performance across MCDM methods. For example, the SAW, TOPSIS, and VIKOR methods were jointly applied to rank banks in Taiwan, with criteria weights determined by AHP. All three methods converged on a unified set of bank rankings [30]. In another instance, VIKOR, TOPSIS, and ELECTRE were used to evaluate three Iranian banks, with criteria weights calculated by AHP. The results indicated that all three methods consistently produced a single set of rankings for the three banks [31].

However, other studies have revealed significant performance discrepancies among different MCDM methods when evaluating banks. For example, ARAS, TOPSIS, and COPRAS were used to assess the financial performance of several banks on Borsa Istanbul (Turkey), with CRITIC employed for criteria weighting. While ARAS, TOPSIS, and COPRAS unanimously identified one bank as having the best financial performance, the rankings of the other banks varied considerably across the different methods [32]. In [33], the RAM, PSI, and SRP methods were collectively applied to evaluate 28 banks in Vietnam. Through Spearman correlation analysis, this study concluded that all three methods RAM, PSI, and SRP were suitable for use. Nonetheless, this research exclusively utilized the Equal method for criteria weighting, which is a clear limitation given that MCDM method performance is known to be heavily dependent on the chosen weighting method [21]. When applied to banks on Borsa Istanbul, the EDAS, MOORA, OCRA, and TOPSIS methods yielded significantly different bank rankings [12]. Furthermore, TOPSIS and GRA were used to evaluate 10 banks in China, 10 in India, 10 in Pakistan, and 8 in Thailand, with criteria weights calculated by AHP. Although both TOPSIS and GRA consistently identified the top-performing bank in each country, the rankings of the remaining banks (within each country) diverged notably between the two methods. This suggests that a comprehensive assessment of the banking system in each nation was not fully achieved [34].

The brief overview of the aforementioned studies underscores the necessity of evaluating banking systems through the concurrent application of multiple MCDM methods and diverse criteria weighting techniques. This approach is essential for identifying the most suitable MCDM and weighting methods, thereby ensuring highly reliable evaluation results for banking systems.

In the current study, three methods Probability, TOPSIS, and RAM are employed concurrently to evaluate the banking system. Simultaneously, five weighting methods Equal, Entropy, MEREC, LOPCOW, and SPC are also utilized for criteria weighting. The rationale for selecting these specific methods was presented in the introduction of this paper.

## 3. Materials and methods

### 3.1. Bank financial performance data

Table 1 consolidates the financial performance data for 19 ban**ks**, designated from B1 to B19. Each bank's performance is characterized by seven specific criteria:
- C1: Loan-to-asset ratio (total loans / total assets)
- C2: Loan-to-deposit ratio (total loans / total deposits)
- C3: Equity-to-asset ratio (total equity / total assets)
- C4: Net profit-to-asset ratio (net periodic profit / total assets)
- C5: Net profit-to-equity ratio (net periodic profit / total equity)
- C6: Branch-to-net profit ratio (number of branches / net periodic profit)
- C7: Employee-to-net profit ratio (number of employees / net periodic profit)

Criteria C1 through C5 are classified as benefit-type criteria (B), meaning higher values are desirable. Conversely, criteria C6 and C7 are cost-type criteria (C), where lower values are preferred [35].

**Table 1**. Bank financial performance data [35]

| Alt. | C1 | C2 | C3 | C4 | C5 | C6 | C7 |
|---|---|---|---|---|---|---|---|
|  | B | B | B | B | B | C | C |
|  | max | | | | | min | |
|  | 0.6363 | 1.1095 | 0.1757 | 0.0558 | 0.4579 | 0.0118 | 0.1301 |
| B1 | 0.5478 | 0.7281 | 0.0876 | 0.0178 | 0.2029 | 0.0428 | 0.5958 |
| B2 | 0.5673 | 0.8457 | 0.0636 | 0.0143 | 0.2245 | 0.0395 | 0.7062 |
| B3 | 0.5556 | 0.8403 | 0.1359 | 0.0437 | 0.3216 | 0.0184 | 0.3788 |
| B4 | 0.6053 | 0.7942 | 0.0645 | 0.0106 | 0.1642 | 0.0704 | 1.4085 |
| B5 | 0.581 | 0.8466 | 0.1325 | 0.0508 | 0.3832 | 0.0143 | 0.3169 |
| B6 | 0.5455 | 0.8984 | 0.1139 | 0.0476 | 0.4178 | 0.0152 | 0.2926 |
| B7 | 0.5212 | 0.8249 | 0.1429 | 0.0558 | 0.3908 | 0.0118 | 0.2119 |
| B8 | 0.6007 | 0.9168 | 0.0736 | 0.0286 | 0.3891 | 0.0253 | 0.6634 |
| B9 | 0.5721 | 0.8514 | 0.1036 | 0.0326 | 0.315 | 0.0391 | 0.7652 |
| B10 | 0.5427 | 0.7452 | 0.0921 | 0.0407 | 0.4416 | 0.04 | 0.7795 |
| B11 | 0.6327 | 0.9294 | 0.1368 | 0.0266 | 0.1943 | 0.0564 | 1.1381 |
| B12 | 0.4423 | 0.5461 | 0.0843 | 0.0342 | 0.4064 | 0.0232 | 0.6126 |
| B13 | 0.5334 | 0.8436 | 0.0879 | 0.036 | 0.4099 | 0.0165 | 0.7408 |
| B14 | 0.4716 | 0.6609 | 0.0731 | 0.0124 | 0.1693 | 0.0563 | 1.3010 |
| B15 | 0.5656 | 0.7648 | 0.0742 | 0.0238 | 0.3199 | 0.1584 | 2.2814 |
| B16 | 0.4436 | 1.1095 | 0.0474 | 0.0217 | 0.4579 | 0.0296 | 0.5734 |
| B17 | 0.6363 | 0.9184 | 0.0881 | 0.0208 | 0.2359 | 0.0266 | 0.8488 |
| B18 | 0.5976 | 1.0592 | 0.0678 | 0.0179 | 0.2635 | 0.0321 | 0.8248 |
| B19 | 0.5463 | 0.7063 | 0.1757 | 0.0488 | 0.2781 | 0.0604 | 0.8705 |

It's observable that the maximum values for benefit-type criteria (C1, C2, C3, C4, C5) are found in banks B17, B16, B19, B7, and B16 respectively. In contrast, the minimum values for cost-type criteria (C6 and C7) are associated with banks B7 and B14. This clearly demonstrates that a simple inspection of Table 1 is insufficient for ranking the financial performance of these banks. Instead, to rank the banks effectively, it is imperative to determine the importance (or weight) **of** each criterion before applying MCDM methods to generate rankings based on these established weights.

## 3.2. Criteria weighting methods

To determine the weights for the various criteria, we first construct a decision matrix. This matrix summarizes the number of alternatives to be ranked and the criteria used to evaluate each alternative. Let *m* represent the number of alternatives to be ranked and *n* denote the number of criteria for each alternative. The value of criterion *j* for alternative *i* is designated as $x_{ij}$, where *i=1…m* and *j=1…n*.

The Equal method calculates criteria weights using the following formula (1):

$$w_j = \frac{1}{n} \tag{1}$$

The Entropy method is applied to compute criteria weights using formulas (2) through (4) sequentially [22, 33].

$$n_{ij} = \frac{y_{ij}}{m + \sum_{i=1}^{m} y_{ij}^2} \tag{2}$$

$$e_j = \sum_{i=1}^{m}\left[n_{ij} \times \ln(n_{ij})\right] - \left(1 - \sum_{i=1}^{m} n_{ij}\right) \times \ln\left(1 - \sum_{i=1}^{m} n_{ij}\right) \tag{3}$$

$$w_j = \frac{1 - e_j}{\sum_{j=1}^{n}(1 - e_j)} \tag{4}$$

For the MEREC method, criteria weights are derived by sequentially applying formulas (5) through (10) [22, 23].

$$n_{ij} = \frac{min\, y_{ij}}{y_{ij}}, \quad if \quad j \in B \tag{5}$$

$$n_{ij} = \frac{y_{ij}}{max\, y_{ij}}, \quad if \quad j \in C \tag{6}$$

$$S_i = Ln\left[1 + \left(\frac{1}{n}\sum_{j}^{n}\left|ln(n_{ij})\right|\right)\right] \tag{7}$$

$$S'_{ij} = Ln\left[1 + \left(\frac{1}{n}\sum_{k,k \neq j}^{n}\left|ln(n_{ij})\right|\right)\right] \tag{8}$$

$$E_j = \sum_{i}^{m}\left|S'_{ij} - S_i\right| \tag{9}$$

$$w_j = \frac{E_j}{\sum_{k}^{n} E_k} \tag{10}$$

The LOPCOW method determines criteria weights by applying formulas (11) through (14) sequentially, where s represents the standard deviation [24].

$$n_{ij} = \frac{x_{ij} - min(x_{ij})}{max(x_{ij}) - min(x_{ij})}, \quad if \quad j \in B \tag{11}$$

$$n_{ij} = \frac{max(x_{ij}) - x_{ij}}{max(x_{ij}) - min(x_{ij})}, \quad if \quad j \in C \tag{12}$$

$$PV_{ij} = \left| \ln \frac{\sqrt{\frac{\sum_{i=1}^{m} r_{ij}^2}{m}}}{\sigma} \right| \cdot 100 \tag{13}$$

$$w_j = \frac{PV_{ij}}{\sum_{j=1}^{n} PV_{ij}} \tag{14}$$

Finally, the SPC method calculates criteria weights by sequentially applying formulas (15) through (19) [25].

$$SPC_j = \frac{\max(x_{ij}) + \min(x_{ij})}{2}; i = 1, 2, \ldots, m; \forall j \in [1 \div n] \tag{15}$$

$$D = |d_{ij}|_{mxn} = \begin{bmatrix} |x_{11} - SPC_1| & |x_{12} - SPC_2| & \cdots & |x_{1n} - SPC_n| \\ |x_{21} - SPC_1| & |x_{22} - SPC_2| & \cdots & |x_{2n} - SPC_n| \\ \cdots & \cdots & \cdots & \cdots \\ |x_{m1} - SPC_1| & |x_{m2} - SPC_2| & \cdots & |x_{mn} - SPC_2| \end{bmatrix} \tag{16}$$

$$R = [r_{ij}]_{m \times n} = \begin{bmatrix} \left|\frac{\sum_{i=1}^{m} d_{i1}}{m \times x_{11}}\right| & \left|\frac{\sum_{i=1}^{m} d_{i2}}{m \times x_{12}}\right| & \cdots & \left|\frac{\sum_{i=1}^{m} d_{in}}{m \times x_{1n}}\right| \\ \left|\frac{\sum_{i=1}^{m} d_{i1}}{m \times x_{21}}\right| & \left|\frac{\sum_{i=1}^{m} d_{i2}}{m \times x_{22}}\right| & \cdots & \left|\frac{\sum_{i=1}^{m} d_{in}}{m \times x_{2n}}\right| \\ \cdots & \cdots & \cdots & \cdots \\ \left|\frac{\sum_{i=1}^{m} d_{i1}}{m \times x_{m1}}\right| & \left|\frac{\sum_{i=1}^{m} d_{i2}}{m \times x_{m2}}\right| & \cdots & \left|\frac{\sum_{i=1}^{m} d_{in}}{m \times x_{mn}}\right| \end{bmatrix} \tag{17}$$

$$Q = [q_{1j}]_{1 \times n} = \left[ \frac{\sum_{i=1}^{m} r_{i1}}{m} \quad \frac{\sum_{i=1}^{m} r_{i2}}{m} \cdots \frac{\sum_{i=1}^{m} r_{in}}{m} \right] \tag{18}$$

$$w_j = [w_{1j}]_{1 \times n} = \left[ \frac{q_1}{\sum_{j=1}^{n} q_j} \quad \frac{q_2}{\sum_{j=1}^{n} q_j} \cdots \frac{q_j}{\sum_{j=1}^{n} q_j} \right] \tag{19}$$

### 3.3. Alternative ranking methods

Following the establishment of the decision matrix as described in Section 3.1, the application of various methods for ranking alternatives is performed as follows:

To rank alternatives using the Probability method, the following sequence of steps is executed [15, 16]:

**Step 1:** Calculate the Probability of Favorable Outcome.

For benefit-type criteria, the probability of a favorable outcome in the decision-making process increases linearly and is calculated using formula (20):

$$P_{ij} \infty x_{ij}, \quad P_{ij} = \alpha_j x_{ij}, i = 1, 2, \ldots, m, \quad j = 1, 2, \ldots n \tag{20}$$

Where $\alpha_j$ is the normalization coefficient for the j-th benefit criterion, calculated as:

$$\alpha_j = \frac{1}{\sum_{i=1}^{m} x_{ij}} \tag{21}$$

For cost-type criteria, the probability of a favorable outcome in the decision-making process is also a linear function and is calculated using formula (22):

$$P_{ij} \infty (x_{jmax} + x_{jmin} - x_{ij}), \quad P_{ij} = \beta_j (x_{jmax} + x_{jmin} - x_{ij}) \\ i = 1, 2, \ldots, m, \quad j = 1, 2, \ldots n \tag{22}$$

Where $\beta_j$ is the normalization coefficient for the j-th cost criterion, calculated as:

$$\beta_j = \frac{1}{m\left(x_{jmax} + x_{jmin} - \frac{\sum_{i=1}^{m} x_{ij}}{m}\right)} \tag{23}$$

**Step 2:** Calculate the Overall Favorable Probability for Each Alternative. This is determined using formula (24):

$$P_i = \prod_{j=1}^{n} (P_{ij})^{w_j} \tag{24}$$

**Step 3:** Alternatives are ranked based on the principle that the best alternative is the one with the highest overall probability.

To rank alternatives using the TOPSIS method, the following steps are applied [13, 14]:

**Step 1:** Calculate the normalized values using formula (25):

$$n_{ij} = \frac{x_{ij}}{\sqrt{\sum_{i=1}^{n} x_{ij}^2}} \tag{25}$$

**Step 2:** Determine the weighted normalized values considering criteria weights using formula (26):

$$y_{ij} = w_j \cdot n_{ij} \tag{26}$$

**Step 3:** Determine the Positive-Ideal Solution ($A^+$) and Negative-Ideal Solution ($A^-$). These are identified for each criterion using formulas (27) and (28), respectively:

$$A^+ = \{y_1^+, y_2^+, \ldots, y_j^+, \ldots, y_n^+\} \tag{27}$$

$$A^- = \{y_1^-, y_2^-, \ldots, y_j^-, \ldots, y_n^-\} \tag{28}$$

Where $y_j^+$ is best and $y_j^-$ is worst are, respectively, the best and worst normalized values of criterion j.

**Step 4:** Determine the values $S_i^+$ and $S_i^-$ using formulas (29) and (30):

$$S_i^+ = \sqrt{\sum_{j=1}^{n}(y_{ij} - y_j^+)^2} \quad i = 1, 2, \ldots, m \tag{29}$$

$$S_i^- = \sqrt{\sum_{j=1}^{n}(y_{ij} - y_j^-)^2} \quad i = 1, 2, \ldots, m \tag{40}$$

**Step 5:** Determine the $C_i$ values using formula (31):

$$C_i = \frac{S_i^-}{S_i^+ + S_i^-} \quad i = 1, 2, \ldots, m; \ 0 \leq C_i^* \leq 1 \tag{31}$$

**Step 6:** Alternatives are ranked based on the principle that the best alternative is the one with the largest $C_i$ value.

The steps for using the RAM method to rank alternatives are as follows [17, 18]:

**Step 1:** Normalize the raw data using formula (32):

$$r_{ij} = \frac{x_{ij}}{\sum_{i=1}^{m} x_{ij}} \tag{32}$$

**Step 2:** Calculate the weighted normalized values for each criterion using formula (33):

$$y_{ij} = w_j \cdot r_{ij} \tag{33}$$

**Step 3:.** Determine the total weighted normalized score using formulas (34) and (35):

$$S_{+i} = \sum_{j=1}^{n} y_{+ij} \quad if \ j \in B \tag{34}$$

$$S_{-i} = \sum_{j=1}^{n} y_{-ij} \quad if \ j \in C \tag{35}$$

**Step 4:** Calculate each alternative's final score using formula (36):

$$RI_i = \sqrt[2+S_{-i}]{2 + S_{+i}} \tag{36}$$

**Step 5:** Rank the alternatives in descending order based on their calculated scores.

### 3.4. Criteria for method comparison

To compare the various MCDM methods and weighting techniques, two primary metrics were employed: the $R_{score}$ coefficient and the Spearman's rank correlation coefficient (hereafter referred to as the Spearman coefficient).

The $R_{score}$ coefficient reflects the stability of alternative rankings when generated by a specific MCDM method. A lower Rscore value indicates greater stability in the rankings. For the Probability, TOPSIS, and RAM methods, this coefficient is calculated using formulas (37), (38), and (39) respectively [36]:

$$R_{score} = \frac{max(P_i)}{min(P_i)}, \quad i = 1 \div m \tag{37}$$

$$R_{score} = \frac{max(C_i)}{min(C_i)}, \quad i = 1 \div m \tag{38}$$

$$R_{score} = \frac{max(RI_i)}{min(RI_i)}, \quad i = 1 \div m \tag{39}$$

The Spearman coefficient, on the other hand, measures the consistency of alternative rankings either when evaluated by different MCDM methods or when evaluated by a single MCDM method but with varying criteria weighting techniques. This coefficient is computed using formula (40) [24, 37]. Where $D_i$ represents the difference in rank for alternative $i$ when compared across different MCDM methods.

$$S = 1 - \frac{6 \sum_{i=1}^{m} D_i^2}{m(m^2 - 1)} \tag{40}$$

### 4. Results and discussion

We calculated the criteria weights using the five distinct methods (formulas 1 to 19), and these results are summarized in Table 2. For clearer commentary, the data from Table 2 is also visually represented in Figure 1.

**Table 2.** Criteria weights

| Method | C1 | C2 | C3 | C4 | C5 | C6 | C7 |
|---|---|---|---|---|---|---|---|
| Equal | 0.1429 | 0.1429 | 0.1429 | 0.1429 | 0.1429 | 0.1429 | 0.1429 |
| Entropy | 0.1809 | 0.1926 | 0.1116 | 0.0920 | 0.1535 | 0.0949 | 0.1745 |
| MEREC | 0.0381 | 0.0722 | 0.1137 | 0.1673 | 0.1061 | 0.2755 | 0.2270 |
| LOPCOW | 0.1562 | 0.0931 | 0.1672 | 0.2321 | 0.1173 | 0.2051 | 0.0289 |
| SPC | 0.0178 | 0.0261 | 0.0824 | 0.1110 | 0.0658 | 0.4228 | 0.2740 |

| | max/min | 10.15 | 7.37 | 2.03 | 2.52 | 2.33 | 4.45 | 9.48 |
|---|---|---|---|---|---|---|---|---|

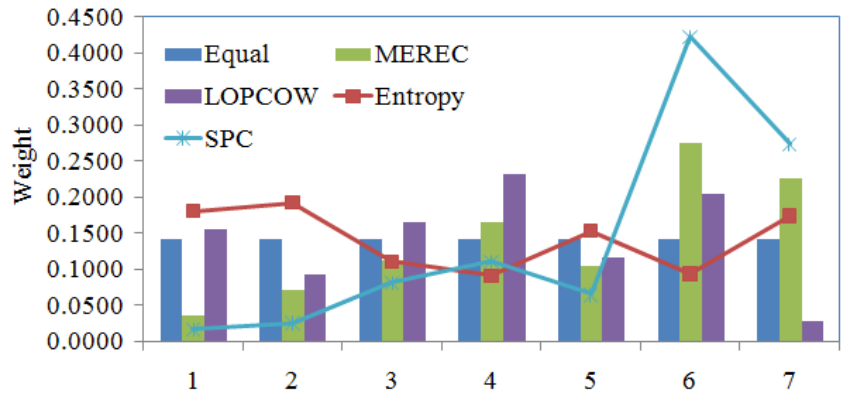

Figure 1. Criteria weights chart

A significant observation is the substantial variation in criteria weights across different calculation methods. For instance, the weight of C1 changed by a factor of 10.15, and C7 by 9.48, while even the least variable criterion, C3, showed a 2.03-fold change. This underscores the critical need to compare and select an appropriate weighting method for specific contexts. Furthermore, among the five weighting methods, we noted a contrasting phenomenon: criteria with small weights using the Entropy method often exhibited large weights when calculated with the SPC method. For example, C1 had the smallest weight with Entropy but the largest with SPC. Conversely, C6 had the largest weight with SPC (0.4228) but the smallest with Entropy (0.0949). This paper will address whether these characteristics lead to significant performance differences between the Entropy and SPC methods.

We applied formulas (20) to (24) to compute the $P_i$ scores for the banks using the Probability method. This process was executed five times, corresponding to the criteria weights derived from each of the five different weighting methods. All computed results are consolidated in Table 3.

Similarly, we applied formulas (25) to (31) to calculate the $C_i$ scores for the banks using the TOPSIS method. This was also performed five times, once for each set of criteria weights. The results are summarized in Table 4.

Finally, formulas (32) to (36) were used to calculate the $RI_i$ scores for the banks with the RAM method. This procedure was likewise repeated five times, aligning with the five different weighting methods. All computed results are presented in Table 5.

Table 3. Pi scores and ranks of alternatives using the Probability method

| Alt. | Equal | | Entropy | | MEREC | | LOPCOW | | SPC | |
|---|---|---|---|---|---|---|---|---|---|---|
| | Pi | rank | Pi | rank | Pi | rank | Pi | rank | Pi | rank |
| B1 | 0.0446 | 15 | 0.0456 | 15 | 0.0455 | 15 | 0.0426 | 15 | 0.0483 | 14 |
| B2 | 0.0428 | 16 | 0.0450 | 16 | 0.0430 | 16 | 0.0398 | 16 | 0.0462 | 15 |
| B3 | 0.0614 | 4 | 0.0596 | 4 | 0.0636 | 4 | 0.0630 | 4 | 0.0637 | 4 |
| B4 | 0.0350 | 18 | 0.0371 | 18 | 0.0325 | 18 | 0.0335 | 18 | 0.0338 | 18 |
| B5 | 0.0651 | 2 | 0.0630 | 2 | 0.0673 | 2 | 0.0673 | 2 | 0.0667 | 2 |
| B6 | 0.0639 | 3 | 0.0625 | 3 | 0.0663 | 3 | 0.0649 | 3 | 0.0658 | 3 |
| B7 | 0.0662 | 1 | 0.0634 | 1 | 0.0698 | 1 | 0.0688 | 1 | 0.0691 | 1 |
| B8 | 0.0542 | 8 | 0.0552 | 6 | 0.0543 | 9 | 0.0530 | 9 | 0.0552 | 7 |
| B9 | 0.0540 | 9 | 0.0538 | 9 | 0.0537 | 10 | 0.0544 | 8 | 0.0534 | 9 |
| B10 | 0.0559 | 7 | 0.0550 | 7 | 0.0562 | 6 | 0.0572 | 6 | 0.0549 | 8 |
| B11 | 0.0494 | 11 | 0.0494 | 13 | 0.0467 | 13 | 0.0508 | 11 | 0.0456 | 17 |
| B12 | 0.0510 | 10 | 0.0494 | 14 | 0.0549 | 8 | 0.0518 | 10 | 0.0568 | 6 |
| B13 | 0.0563 | 6 | 0.0557 | 5 | 0.0576 | 5 | 0.0571 | 7 | 0.0581 | 5 |
| B14 | 0.0394 | 17 | 0.0413 | 17 | 0.0415 | 17 | 0.0355 | 17 | 0.0457 | 16 |
| B15 | 0.0239 | 19 | 0.0252 | 19 | 0.0141 | 19 | 0.0269 | 19 | 0.0090 | 19 |
| B16 | 0.0494 | 12 | 0.0519 | 10 | 0.0504 | 11 | 0.0455 | 13 | 0.0522 | 11 |
| B17 | 0.0490 | 13 | 0.0501 | 11 | 0.0485 | 12 | 0.0481 | 12 | 0.0506 | 12 |
| B18 | 0.0473 | 14 | 0.0496 | 12 | 0.0465 | 14 | 0.0448 | 14 | 0.0486 | 13 |
| B19 | 0.0566 | 5 | 0.0540 | 8 | 0.0557 | 7 | 0.0604 | 5 | 0.0524 | 10 |

Table 4. $C_i$ scores and ranks of alternatives using the TOPSIS method

| Alt. | Equal | | Entropy | | MEREC | | LOPCOW | | SPC | |
|---|---|---|---|---|---|---|---|---|---|---|
| | Ci | rank | Ci | rank | Ci | rank | Ci | rank | Ci | rank |
| B1 | 0.6151 | 14 | 0.6329 | 13 | 0.7101 | 13 | 0.580 | 15 | 0.7662 | 13 |
| B2 | 0.5940 | 16 | 0.6153 | 16 | 0.6956 | 14 | 0.563 | 16 | 0.7641 | 14 |
| B3 | 0.8204 | 4 | 0.8037 | 4 | 0.8872 | 4 | 0.832 | 4 | 0.9208 | 4 |
| B4 | 0.4267 | 18 | 0.4089 | 18 | 0.4947 | 18 | 0.443 | 18 | 0.5418 | 18 |
| B5 | 0.8603 | 2 | 0.8431 | 2 | 0.9213 | 2 | 0.880 | 2 | 0.9469 | 2 |
| B6 | 0.8378 | 3 | 0.8385 | 3 | 0.9085 | 3 | 0.840 | 3 | 0.9421 | 3 |
| B7 | 0.8802 | 1 | 0.8545 | 1 | 0.9478 | 1 | 0.902 | 1 | 0.9714 | 1 |
| B8 | 0.6920 | 8 | 0.7042 | 6 | 0.7779 | 7 | 0.681 | 10 | 0.8365 | 7 |
| B9 | 0.6804 | 9 | 0.6732 | 10 | 0.7406 | 10 | 0.685 | 9 | 0.7753 | 11 |
| B10 | 0.6939 | 7 | 0.6778 | 8 | 0.7477 | 9 | 0.715 | 6 | 0.7725 | 12 |
| B11 | 0.5786 | 17 | 0.5525 | 17 | 0.6162 | 17 | 0.605 | 13 | 0.6479 | 17 |
| B12 | 0.6981 | 6 | 0.6740 | 9 | 0.8022 | 6 | 0.710 | 7 | 0.8560 | 5 |
| B13 | 0.7245 | 5 | 0.7062 | 5 | 0.8023 | 5 | 0.745 | 5 | 0.8525 | 6 |
| B14 | 0.5962 | 15 | 0.6472 | 11 | 0.6864 | 15 | 0.504 | 17 | 0.7261 | 15 |
| B15 | 0.1574 | 19 | 0.1679 | 19 | 0.0847 | 19 | 0.159 | 19 | 0.0407 | 19 |
| B16 | 0.6562 | 11 | 0.6917 | 7 | 0.7547 | 8 | 0.618 | 12 | 0.8238 | 8 |
| B17 | 0.6397 | 12 | 0.6322 | 14 | 0.7306 | 11 | 0.642 | 11 | 0.7995 | 9 |
| B18 | 0.6188 | 13 | 0.6307 | 15 | 0.7116 | 12 | 0.605 | 14 | 0.7814 | 10 |
| B19 | 0.6723 | 10 | 0.6400 | 12 | 0.6730 | 16 | 0.706 | 8 | 0.6680 | 16 |

Table 5. $RI_i$ scores and ranks of alternatives using the RAM method

| Alt. | Equal | | Entropy | | MEREC | | LOPCOW | | SPC | |
|---|---|---|---|---|---|---|---|---|---|---|
| | RIi | rank | RIi | rank | RIi | rank | RIi | rank | RIi | rank |
| B1 | 1.4213 | 15 | 1.4222 | 15 | 1.4150 | 14 | 1.4221 | 15 | 1.4099 | 14 |
| B2 | 1.4209 | 16 | 1.4221 | 16 | 1.4143 | 16 | 1.4215 | 16 | 1.4093 | 16 |
| B3 | 1.4279 | 4 | 1.4276 | 4 | 1.4225 | 4 | 1.4300 | 4 | 1.4171 | 4 |
| B4 | 1.4170 | 18 | 1.4183 | 18 | 1.4082 | 18 | 1.4184 | 18 | 1.4018 | 18 |
| B5 | 1.4294 | 2 | 1.4290 | 2 | 1.4241 | 2 | 1.4318 | 2 | 1.4186 | 2 |
| B6 | 1.4289 | 3 | 1.4288 | 3 | 1.4237 | 3 | 1.4309 | 3 | 1.4182 | 3 |
| B7 | 1.4302 | 1 | 1.4295 | 1 | 1.4254 | 1 | 1.4327 | 1 | 1.4199 | 1 |
| B8 | 1.4249 | 8 | 1.4256 | 6 | 1.4185 | 7 | 1.4262 | 8 | 1.4132 | 7 |
| B9 | 1.4242 | 9 | 1.4246 | 10 | 1.4173 | 11 | 1.4260 | 10 | 1.4113 | 11 |
| B10 | 1.4250 | 7 | 1.4252 | 8 | 1.4183 | 8 | 1.4272 | 7 | 1.4118 | 9 |
| B11 | 1.4223 | 14 | 1.4229 | 14 | 1.4140 | 17 | 1.4245 | 12 | 1.4071 | 17 |
| B12 | 1.4241 | 10 | 1.4240 | 11 | 1.4190 | 6 | 1.4261 | 9 | 1.4140 | 6 |
| B13 | 1.4257 | 5 | 1.4258 | 5 | 1.4199 | 5 | 1.4279 | 6 | 1.4147 | 5 |
| B14 | 1.4202 | 17 | 1.4216 | 17 | 1.4144 | 15 | 1.4195 | 17 | 1.4095 | 15 |
| B15 | 1.4133 | 19 | 1.4150 | 19 | 1.4001 | 19 | 1.4152 | 19 | 1.3889 | 19 |
| B16 | 1.4240 | 11 | 1.4253 | 7 | 1.4178 | 9 | 1.4243 | 13 | 1.4124 | 8 |
| B17 | 1.4230 | 12 | 1.4237 | 12 | 1.4163 | 12 | 1.4246 | 11 | 1.4113 | 10 |
| B18 | 1.4225 | 13 | 1.4236 | 13 | 1.4155 | 13 | 1.4235 | 14 | 1.4103 | 12 |
| B19 | 1.4253 | 6 | 1.4249 | 9 | 1.4177 | 10 | 1.4283 | 5 | 1.4100 | 13 |

An examination of Table 3 reveals that when the Probability method is employed for ranking banks, the ranks exhibit minimal changes despite using five different sets of criteria weights. It's noteworthy, as shown in Table 2, that the criteria weights themselves varied considerably. Despite this, consistently top-ranked banks were identified regardless of the weighting method used. Specifically, B7 consistently ranked 1st, B5 2nd, B6 3rd, and B3 4th. Furthermore, the lowest-ranked banks, B15 (19th) and B4 (18th), also remained consistent across all five weighting methods. This partially demonstrates the Probability method's advantage in ensuring ranking stability when various weighting methods are applied to criteria.

Table 4 also shows that the top-ranked banks remain consistent when using the TOPSIS method, even with varying criteria weights. Specifically, TOPSIS consistently identified B7 as 1st, B5 as 2nd, B6 as 3rd, and B3 as 4th across all five weighting methods. Similarly, the two lowest-ranked banks, B15 (19th) and B4 (18th), were also consistently identified. A notable observation is that the four best-performing and two worst-performing banks were consistently identified across both the Probability and TOPSIS methods.

When applying the RAM method to rank alternatives, the top-ranked banks B7 (1st), B5 (2nd), B6 (3rd), and B3 (4th) were consistently identified, irrespective of the weighting method used. The two lowest-ranked banks, B4 (18th) and B15 (19th), also showed consistent rankings across different weighting methods. A thorough review of Tables 3, 4, and 5 reveals a highly significant finding: the best and worst-performing banks consistently maintained their ranks, irrespective of the criteria weighting method or the alternative ranking method employed. Specifically, B7 consistently ranked 1st, B5 2nd, B6 3rd, B3 4th, B15 19th, and B4 18th across all methods.

Therefore, in terms of identifying the highest and lowest-ranked banks, the Equal, Entropy, MEREC, LOPCOW, and SPC weighting methods exhibit comparable performance. Similarly, the Probability, TOPSIS, and RAM alternative ranking methods are also deemed to have equivalent performance in this regard. However, when considering the overall ranking of all banks, numerous rank changes are observed when different weighting methods or alternative ranking methods are used. Hence, to ensure the stability of bank rankings, it is crucial to conduct comparative analyses among the criteria weighting methods and among the alternative ranking methods.

Applying formulas (37) to (39), the $R_{score}$ coefficients were calculated for all performed scenarios, as summarized in Table 6.

**Table 6.** $R_{score}$ coefficient values

| Method | $R_{score}$ | | | | |
|---|---|---|---|---|---|
| | Equal | Entropy | MEREC | LOPCOW | SPC |
| Probability | 2.7678 | **2.5141** | 4.9441 | 2.5588 | 7.6535 |
| TOPSIS | 5.5911 | 5.0883 | 11.1875 | 5.6596 | 23.8572 |
| RAM | 1.0119 | 1.0103 | 1.0181 | 1.0124 | 1.0223 |

It is evident that when the Entropy method is used for criteria weighting, the $R_{score}$ coefficient consistently exhibits the lowest value compared to other weighting methods, regardless of whether Probability, TOPSIS, or RAM is used for ranking. Conversely, employing the SPC method for criteria weighting consistently results in the highest $R$score coefficient, irrespective of the ranking method. This outcome indicates that among the five criteria weighting methods examined, the Entropy method provides the highest stability in alternative rankings. In contrast, the SPC method yields the lowest ranking stability. This is not a critique of the SPC method but rather a recommendation that, among the Equal, Entropy, MEREC, LOPCOW, and SPC weighting methods, Entropy is advised to ensure high stability in alternative rankings.

We applied formula (40) to calculate the Spearman coefficients between the weighting methods for each case using the Probability, TOPSIS, and RAM methods for ranking alternatives. These results are compiled in Tables 7 to 9.

**Table 7.** Spearman correlation coefficients using the Probability method for ranking alternatives

| Method | Entropy | MEREC | LOPCOW | SPC |
|---|---|---|---|---|
| Equal | 0.9596 | 0.9842 | 0.9947 | 0.9246 |
| Entropy | | 0.9526 | 0.9491 | 0.9193 |
| MEREC | | | 0.9789 | 0.9632 |
| LOPCOW | | | | 0.9140 |
| Average | 0.9699 | | | |

**Table 8.** Spearman correlation coefficients using the TOPSIS method for ranking alternatives

| Method | Entropy | MEREC | LOPCOW | SPC |
|---|---|---|---|---|
| Equal | 0.9474 | 0.9491 | 0.9702 | 0.9123 |
| Entropy | | 0.9421 | 0.8842 | 0.8930 |
| MEREC | | | 0.8842 | 0.9825 |
| LOPCOW | | | | 0.8404 |
| Average | 0.9205 | | | |

**Table 9.** Spearman correlation coefficients using the RAM method for ranking alternatives

| Method | Entropy | MEREC | LOPCOW | SPC |
|---|---|---|---|---|
| Equal | 0.9719 | 0.9509 | 0.9877 | 0.9105 |
| Entropy | | 0.9596 | 0.9404 | 0.9439 |

| MEREC | | | 0.9246 | 0.9860 |
|---|---|---|---|---|
| LOPCOW | | | | 0.8772 |
| Average | | 0.9453 | | |

Across Tables 7, 8, and 9, all Spearman coefficients between any two weighting methods are notably high, exceeding 0.8. This suggests that all sets of bank rankings are considered appropriate [38]. However, a more detailed analysis reveals that the Spearman coefficient between any two weighting methods consistently performs better when the Probability method is used for ranking compared to TOPSIS and RAM. When the Probability method ranks alternatives, the minimum Spearman coefficient is 0.9140 (between LOPCOW and SPC). For TOPSIS, the minimum is 0.8404 (also between LOPCOW and SPC), and for RAM, it's 0.8772 (again, between LOPCOW and SPC). This partially suggests that the highest level of ranking stability for alternatives is achieved when the Probability method is used. Conversely, alternative rankings show the weakest stability when the TOPSIS method is employed. This sensitivity of TOPSIS to criteria weights has also been revealed in a recent study [39]. Furthermore, the average Spearman coefficient among weighting methods is 0.9699 when using the Probability method for ranking, 0.9205 when using the TOPSIS method, and 0.9453 when using the RAM method. This further reinforces the conclusion that the highest level of ranking stability is ensured by the Probability method, followed by RAM, and then TOPSIS.

**Conclusion**

Evaluating the financial performance of banks is a critical and intricate task. Achieving highly stable rankings of bank financial efficiency not only generates precise data on banks, fostering healthy competition among them, but also provides transparent information for the public and social organizations, and valuable insights for bank management. The financial performance ranking of banks is heavily influenced by criteria weights and the specific methods employed for ranking. This study aimed to compare five criteria weighting methods Equal, Entropy, MEREC, LOPCOW, and SPC as well as three alternative ranking methods Probability, TOPSIS, and RAM. Several key conclusions can be drawn:

- ✓ A notable inverse relationship exists between the Entropy and SPC weighting methods: if a criterion's weight is low when calculated by Entropy, it tends to be high when calculated by SPC, and vice versa.

- ✓ The top-performing banks (ranks 1 to 4) and the lowest-performing banks (ranks 18 and 19) consistently maintain their positions, regardless of the criteria weighting method or the alternative ranking method utilized.

- ✓ When any of the ranking methods (Probability, TOPSIS, or RAM) are employed, the stability of alternative rankings is maximized when the Entropy method is used for criteria weighting. Conversely, utilizing the SPC method for criteria weighting leads to the least stable alternative rankings. This suggests a strong recommendation for using the Entropy method to determine criteria weights.

- ✓ When different criteria weighting methods are applied, the stability of alternative rankings is highest when the Probability method is used for ranking. In contrast, the TOPSIS method yields the lowest ranking stability. This implies that the Probability method should be favored for ranking alternatives.

- A limitation of this study is the exclusive use of objective weighting methods (Equal, Entropy, MEREC, LOPCOW, and SPC) to calculate criteria weights. These methods do not incorporate the opinions of financial experts regarding the importance of different criteria. A more practical assessment of bank financial performance could potentially be achieved by considering expert opinions on criteria significance. Future research could explore subjective weighting methods that integrate expert judgment, such as SIWEC, PIPRECIA, ROC, etc., to achieve a more comprehensive evaluation of banks.